\documentclass[letterpaper, 10 pt, conference]{ieeeconf}  % Comment this line out if you need a4paper

\IEEEoverridecommandlockouts                              % This command is only needed if 
\usepackage{graphicx} % for pdf, bitmapped graphics files
\usepackage{times} % assumes new font selection scheme installed
\usepackage{amsmath} % assumes amsmath package installed
\usepackage{textpos}
\usepackage{myMacros}
\usepackage{blindtext}
\usepackage{booktabs}
\usepackage{algorithm}
\usepackage{array}
\DeclareMathOperator*{\argmax}{arg\,max}
\DeclareMathOperator*{\argmin}{arg\,min}

\title{\LARGE \bf Joint Constrained Bayesian Optimization of Planning, Guidance, Control, and State Estimation of an 
Autonomous Underwater Vehicle$^\dagger$}

\author{David Stenger$^{1}$, Maximilian Nitsch$^{1}$ and Dirk Abel$^{1}$% <-this % stops a space
\thanks{$^\dagger$ This publication results from the project line "Technologies for Rapid Ice Penetration and subglacial Lake Exploration" (TRIPLE) and is supported
by the German Federal Ministry for Economic Affairs
and Energy (grant 50NA2009). Basis for the support is a decision by the German Bundestag.}
\thanks{$^{1}$ Institute of Automatic Control,
        RWTH Aachen University, Aachen, Germany
        {\tt\small \{d.stenger@irt.rwth-aachen.de\}}}
}

\begin{document}

\maketitle
\thispagestyle{empty}
\pagestyle{empty}

%%%%%%%%%%%%%%%%%%%%%%%%%%%%%%%%%%%%%%%%%%%%%%%%%%%%%%%%%%%%%%%%%%%%%%%%%%%%%%%%
\begin{abstract}

The performance of a 
guidance, navigation and control (GNC) system of an autonomous underwater vehicle (AUV) 
heavily depends on the correct tuning of its 
parameters.  
Our objective is to automatically tune these parameters with respect to arbitrary high-level control objectives within different operational scenarios.  
In contrast to literature, an overall tuning is performed for the entire GNC system, which is new in the context of autonomous underwater vehicles. The main challenges in solving the optimization problem are computationally expensive objective function evaluations,  
crashing simulations due to infeasible parametrization and the numerous tunable parameters (in our case 13). These challenges are met by using constrained Bayesian optimization with crash constraints.  
The method is demonstrated in simulation on a GNC system of an underactuated miniature AUV designed within the TRIPLE-nanoAUV initiative 
for exploration of sub-glacial lakes.  
We quantify the substantial reduction in energy consumption achieved by tuning the overall system. Furthermore, different parametrizations are automatically generated for different power consumption functions, robustness, and accuracy requirements. E.g. energy consumption can be reduced by $\sim 28 \%$, if the maximum allowed deviation from the planned path is increased by $\sim 65 \%$. This shows the versatile practical applicability of the optimization-based tuning approach.

\end{abstract}

\begin{textblock*}{\textwidth}(0mm,-130mm)
	\small\textcopyright 2022 IEEE. Personal use of this material is permitted. Permission from IEEE must be obtained for all other uses, in any
	current or future media, including reprinting/republishing this material for advertising or promotional purposes, creating
	new collective works, for resale or redistribution to servers or lists, or reuse of any copyrighted component of this work in
	other works.
\end{textblock*}
%%%%%%%%%%%%%%%%%%%%%%%%%%%%%%%%%%%%%%%%%%%%%%%%%%%%%%%%%%%%%%%%%%%%%%%%%%%%%%%%
\section{INTRODUCTION}

Successful operation of an autonomous underwater vehicle (AUV) requires the close collaboration of state estimation, guidance, control, and path planning algorithms within a guidance navigation and control (GNC) framework. For each of those domains appropriate algorithms were established in simulation and real world experiments 
e.g. \cite{Xiang.2018}, \cite{Zeng.2015},  \cite{Melo.2017}.  

However, the performance of the overall system heavily depends on the correct tuning of their  
algorithmic parameters for the specific task and vehicle at hand. These parameters govern e.g. the aggressiveness of the guidance algorithm.  
For some GNC algorithms, analytical tuning methods exist.  
However, they can only be applied to a limited range of scenarios and control objectives (e.g. LQR).  
In contrast, we focus on the tuning of the overall GNC system and  
allow arbitrary high-level objective functions and constraints. 

The tuning problem is formulated as a \textbf{black-box optimization problem} to be approximately solved  
in simulation.
Solving black-box problems is challenging due to the unknown analytic relationship between the GNC parameters and the obtained objective function values. 
Additionally, poor parametrization may result e.g. in diverging state estimation solutions leading to simulation crash. This setting is known as learning under crash constraints (LCC) \cite{Marco.2021}. Furthermore,  closed-loop simulations are expensive in terms of CPU time. Thus, sample-efficiency of the used optimizer is important.

In this contribution, we use constrained \textbf{Bayesian optimization (BO)} with crash constraints to meet these challenges. BO has become a common method used for sample-efficient black-box optimization e.g. in control \cite{Marco.2016}  
and state estimation \cite{Gehrt.1123202011242020}. Constrained BO in control engineering \cite{ChristopherKonig.2020} as well as LCC \cite{Marco.2021} have previously been addressed.  

In the context of \textbf{automatic tuning for AUVs}, to the best of the authors knowledge, tuning of the overall GNC system has not been considered yet. Different metaheuristics  
have been applied for PID tuning for AUVs  
\cite{Herlambang.2019}. However, such metaheuristics were shown to be less sample-efficient than BO (cf. e.g. \cite{Acerbi.2017, Dorschel.2021}). 
BO has been used to fairly compare different controllers  
for AUVs \cite{M.Manhaes.2017}. Simplified  
 optimal state knowledge was assumed, though. 
The tuning of parameters of navigation filters for AUVs has only been reported using partical swarm optimization \cite{8962573}.
E.g. in  
\cite{Gehrt.1123202011242020}, the potential of BO for filter tuning has been show for a different application. However, in those publications 
the effect of filter design on the control performance was not considered.

In this contribution, we showcase the BO-based tuning method  
on an underactuated AUV designed for the exploration of subglacial lakes and oceans within the \textbf{TRIPLE-nanoAUV} initiative \cite{Waldmann2020, Nitsch2021}. For this purpose, a GNC system consisting of Dubins paths for planning, a side slip compensating line of sight guidance law, a LQR controlling different thrusters, and an UKF for state and current estimation is developed.  
In total 13 GNC parameters spanning all modules are optimized. This paper uses tuning in simulation to evaluate the versatility of the GNC system and find initial parameters for experiments with different accuracy requirements.  
Therefore preventing constraint violations during optimization by using safe BO (e.g. \cite{Sui.2015}) is not required.  
  
Three main results are presented: Each component of the GNC system has a significant influence on overall system performance. Different accuracy requirements and power consumption function can be facilitated with the same policy structure. Robustness can be increased by tuning on a variety of operational scenarios.

This paper is structured as follows: First, in Sec.\,\ref{sec:parameter_tuning_method} the parameter tuning method is introduced. Afterwards, the nanoAUV's model (cf. Sec.\, \ref{sec:auv_model}) and its GNC system (cf. Sec.\, \ref{sec:gnc_system}) are presented. Sec.\, \ref{sec:simulation_scenario} outlines the considered simulation scenario, and results of the optimization are given in Sec.\,\ref{sec:results}. 

\section{PARAMETER TUNING METHOD} \label{sec:parameter_tuning_method}

\subsection{Optimization problem formulation}

\begin{figure}[H]
    \centering
    \includegraphics[width=0.8\linewidth]{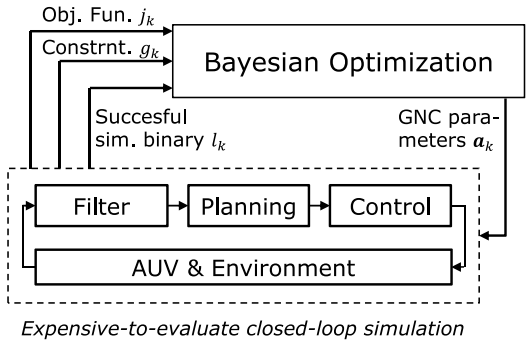}
    \caption{Considered black-box optimization setting.}
    \label{fig:BBOverview}
\end{figure}

Fig. \ref{fig:BBOverview} gives an overview over the black-box optimization setting we consider in this contribution. Let $\mathbf{a}$ denote the parameters of the various modules of the GNC system. At each iteration $k$ of the optimization, the closed loop of GNC system with state estimation (filter), planning, guidance, and control as well as the AUV in its environment is simulated with specific GNC parameters $\mathbf{a}_k$ (cf. Sec. \ref{sec:gnc_system}). Within the simulation the AUV fulfills a set of specific task, in our case following a reference path. The results of the simulation depend on the parameter combination $\mathbf{a}_k$ which is queried by the optimizer. The value of the cost function $j_k$, in our case total energy consumption, and constraint value $g_k$, in our case the maximum deviation from the reference path, are returned to the optimizer. Additionally, a binary value $l_k$ is returned which indicates whether the simulation was successful ($l_k = 1$) or not ($l_k = 0$). Reasons for a simulation not being successful are for example diverging state estimation solutions resulting in the AUV not reaching its desired goal state. In this case, returned $j_k$ and $g_k$ values are not meaningful.  

With these definitions we aim at solving the following deterministic constraint optimization problem:

\begin{align} \label{eq:generalOptProb}
  \quad \boldsymbol{a}^\star =  \argmin \quad \quad \quad &J(\boldsymbol{a}) \\
\text{s.t.} \, \, \, \, \, \, \, \quad \quad   \boldsymbol{a}_{\mathrm{min}} &\leq \boldsymbol{a} \leq \boldsymbol{a}_{\mathrm{max}} \nonumber  \\ 
  g(\boldsymbol{a}) &\leq g_{\mathrm{max}} \nonumber \\ 
  l(\boldsymbol{a}) &= 1 \nonumber 
\end{align}

Note that the optimization problem is deterministic, because the same draw of disturbance, and model plant mismatch (cf. Sec. \ref{sec:simulation_scenario}) is used during optimization. In Section \ref{resultsD} robust optimization is performed on the identical five realizations and therefore still deterministic.

\subsection{Constrained Bayesian optimization (BO) with max-value entropy search and crash constraints}

The problem stated in \eqref{eq:generalOptProb} is a real-valued and binary constrained deterministic black-box optimization problem with an expensive-to-evaluate objective function. BO has become a standard tool to solve such problems in the context of control. It can be attributed to the \emph{Micro Data RL} branch of Reinforcement Learning, see \cite{Chatzilygeroudis.2020}. Algorithm 1 gives an overview over BO. For a comprehensive introduction to BO the reader is referred to e.g. \cite{Shahriari.2016}. 

\begin{algorithm}[h] 
	1: Initial sampling of $A_{1}$, $J_{1}$, $G_{1}$ and $L_{1}$:  \\ [3pt]
	2: \textbf{for} k = 1; 2; . . . ; \textbf{do} \\[3pt]
	3: \quad $\tilde{J}_{k},\tilde{G}_{k} \leftarrow \textrm{addArtificialData}(A_{k},J_{k},G_{k},L_{k}$) \\[3pt]
	4: \quad update probabilistic surrogate models using  \\
	\hspace*{6.5mm} $A_{k}$, $\tilde{J}_{k}$, and $\tilde{G}_{k}$\\[3pt]
	5: \quad select $\mathbf{a}_{k+1}$ by optimizing an acquisition function:\\ 
	\hspace*{6.5mm} $\mathbf{a}_{k+1} = \argmax_\mathbf{a}(\alpha(\mathbf{a}|A_{k}, \tilde{J}_{k}, \tilde{G}_{k}))$\\[3pt]	
	6: \quad query objective function to obtain $j_{k+1}$, $g_{k+1}$ and $l_{k+1}$ \\[3pt]
	7: \quad augment $A_{k+1} = \{ A_{k}, \mathbf{a}_{k+1}\}$, $J_{k+1} = \{ J_{k},j_{k+1}\}$ \\
	\hspace*{6.5mm}  $G_{k+1} = \{ G_{k},y_{k+1}\}$, $L_{k+1} = \{ L_{k},l_{k+1}\}$ \\[3pt]
	8: \textbf{end for} 
	
	\caption{Bayesian optimization with crash constraints}
	\label{Algo:BayesOpt}
\end{algorithm} 

At iteration $k = 1$, an initial set of random parametrizations $A_{1}$ is generated to obtain initial responses for objective function $J_{1}$, constraint $G_{1}$ and crash constraint $L_{1}$ (cf. Step 1).    
We use Gaussian process regression (GPR) \cite{Rasmussen.2006} as the surrogate model. The \textbf{surrogate model} uses past evaluations  
in order to give a probabilistic estimate of the objective function value and the constraint function at unknown locations (cf.  Step 4). A separate GP model is created for each response with inputs $\{ A_{k}, \tilde{J}_{k} \}$ and $\{ A_{k}, \tilde{G}_{k} \}$, respectively. We use a squared exponential kernel with automated relevance detection and a constant mean function for both models. Since the objective function evaluations are deterministic the noise hyperparameter of the GP model is set to zero. 
Additionally, a smooth box hyperprior is placed on the kernel length scales in order to avoid very large or very small length scales. The GPs hyperparameters are tuned in each iteration by maximizing the a posteriori likelihood via a combination of random search and gradient decent. 

Failed objective function evaluations pose a  
challenge for BO, because in that case the obtained $j$ and $g$ are not meaningful. This setting is known as \textbf{learning under crash constraints} \cite{Marco.2021}. Setting $j$ and $g$ to some arbitrary large value is not a valid option because this may result in a discontinuous objective function landscapes which contradicts the smoothness assumptions encoded in the Gaussian kernel (cf. e.g. \cite{Dorschel.2021} ).  
Instead, here we generate artificial $\tilde{j}$ and $\tilde{g}$ values for the failed simulations (cf. Step 3),
\begin{equation}
    \tilde{j} = \mu_j + 3 \sigma_j \quad \quad \tilde{g} = \mu_g ,     
\end{equation}
where $\mu_j, \mu_g, \sigma_j$ are the mean and standard deviation of the GP predictions generated ignoring the failed simulations.  
This approach is similar to the idea in \cite{Marco.2021}, in that it modifies the GP-posterior in a way such it decreases the probability to observe the optimum around the failed observation 
while not contradicting the smoothness assumptions of the GP.

Constrained max-value entropy search (cMES) \cite{Perrone.15.10.2019} is used as the \textbf{acquisition function} $\alpha$ because it was shown to outperform the more classical expected improvement with constraints (EIC). cMES chooses the next sample point $\mathbf{a}_{k+1}$, such that the mutual information between the evaluation and the estimated distribution of the optimum is maximized, and is therefore considered to be an information theoretic acquisition function (cf. step 6).

\section{The TRIPLE nanoAUV} \label{sec:auv_model}
 
The nanoAUV \cite{Waldmann2020, Nitsch2021} is developed to explore the liquid water column of subglacial lakes. An ice-melting probe transports the nanoAUV to the ice-water interface \cite{heinen2021}. The nanoAUV acts as payload, therefore installation space is very limited which restricts the GNC system. 
As a result, only low performance on-board computers are applicable and the GNC system needs to be developed as simple and computationally efficient as possible.

\subsection{Hydrodynamic model}

We follow Fossen et al. \cite{Fossen.2006} in modeling the AUV's 6 degrees of freedom:
\begin{align} \label{eq:Fossen_Mdl}
    \mathbf{M}_{RB} \dot{\boldsymbol{\nu}} &+ \mathbf{C}_{RB}(\boldsymbol{\nu})\boldsymbol{\nu} + \mathbf{M}_A \dot{\boldsymbol{\nu}}_r + \mathbf{C}_A(\boldsymbol{\nu}_r) \nu_r + \dots \\
    & \mathbf{D}(\boldsymbol{\nu}_r) \boldsymbol{\nu}_r+ \mathbf{g}(\boldsymbol{\eta}) = \boldsymbol{\tau} \ , \nonumber
\end{align} 
with gravity $\mathbf{g}(\boldsymbol{\eta})$, mass matrices $\mathbf{M}_{RB}$, and $\mathbf{M}_{A}$ and coriolis-centripetal matrices $\mathbf{C}_{RB}$, and $\mathbf{C}_{A}$ for the rigid body (RB) and added mass (A). The state vector $\mathbf{x} = [ \boldsymbol{\nu}, \boldsymbol{\eta} ]$ consists of the motion components $\boldsymbol{\nu} = [u,v,w,p,q,r]$ in body frame and $\boldsymbol{\eta} =[n,e,d,\phi, \theta, \psi]$ denotes the positions $n,e,d$ of the body frame in the earth fixed frame. The Euler angles are roll $\phi$, pitch $\theta$ and yaw $\psi$. 
The relative velocity \hbox{$\boldsymbol{\nu}_r = [u-u_c^b, v - v_c^b, w-w_c^b,p,q,r]^\mathrm{T}$} is calculated by considering the current vector $[u_c^b, v_c^b, w_c^b]^\mathrm{T}$.   
We use a linear hydrodynamic model without cross couplings and identical drag in sway/heave as well as pitch/yaw movement:  $\mathbf{D} = \mathrm{diag}[D_1, D_2, D_2,D_4,D_5,D_5]$.  

Because the simulation is considered as a black-box within the optimization, more advanced experimentally validated models can also be used without altering the optimization methodology.

\subsection{Sensors  \& actuators}

The navigation system mainly relies on measurements provided by an  
IMU, a multi-magnetometer  
for attitude estimation and an USBL system  
for positioning. The depth estimation is aided by a 
pressure sensor. The IMU and magnetometer data are fused within an attitude heading reference system (AHRS) algorithm. The AHRS is not optimized and therefore not discussed in detail in this paper. 

Five thrusters are used as the main actuators. 
The surge thruster with control input $u_{\mathrm{surge}}$ is located at the rear of the vehicle, generating a forward pointing force. An additional 4 thrusters are located at the back of the vehicle providing force in the sway ($u_{\mathrm{rl}}$) and heave ($u_{\mathrm{ud}})$ directions. All thrusters except for the surge thruster introduce moments around the center of the vehicle. The thrusters are modeled using a first order lag element. The generalised forces $\boldsymbol{\tau}$ of Eq. \eqref{eq:Fossen_Mdl} are calculated by combining the geometric properties of the thrusters with the respective control signals. Note that this actuator setup results in an underactuated AUV. Additionally, a buoyancy engine as well as a movable mass for pitch motion are used, influencing $\mathbf{M}_{RB}$ and $\mathbf{C}_{RB}$. They are controlled in a feed-forward manner. Since this feed-forward controller is not optimized, it is not described any further.

\section{GNC SYSTEM} \label{sec:gnc_system}

\subsection{State estimation} \label{sec:stateEstimation}
The state estimator is realized as classical UKF, proposed by \cite{wan2000}. The nonlinear state space model is given by:
\begin{align}
\myFrameVecDot{x}{}{} =& ~\myFrameVec{f}{}{}(\myFrameVec{x}{}{},\myFrameVec{u}{}{})+ \myFrameVec{G}{}{}\myFrameVec{w}{}{}  &\myFrameVec{w}{}{} \sim \mathcal{N}(\myFrameVec{0}{}{},\,\myFrameVec{Q}{}{}) \label{eq:transition}\\
\myFrameVec{z}{}{}  =&  ~\myFrameVec{h}{}{}(\myFrameVec{x}{}{}) + \myFrameVec{v}{}{} &\myFrameVec{v}{}{} \sim \mathcal{N}(\myFrameVec{0}{}{},\,\myFrameVec{R}{}{}) \label{eq:measure}
\end{align}
As prediction model $\myFrameVec{f}{}{}(\myFrameVec{x}{}{},\myFrameVec{u}{}{})$ the Fossen model from Eq.\,\eqref{eq:Fossen_Mdl} is used. The actuator control commands act as input $\myFrameVec{u}{}{}$. Furthermore, the surge and sway velocities of the current $\myFrameScalar{u}{c}{b}$ and $\myFrameScalar{v}{c}{b}$ are estimated as random walk. The heave component is neglected. This results in the following process model:
\begin{align}
    \pVecFour{\myFrameVecDot{\nu}{}{}}{\myFrameVecDot{\eta}{}{}}{\myFrameScalarDot{u}{c}{b}}{\myFrameScalarDot{v}{c}{b}} = \pVecFour{\text{Fossen from Eq.\,\eqref{eq:Fossen_Mdl}}}{\text{Fossen from Eq.\,\eqref{eq:Fossen_Mdl}}}{0}{0} + \pVecFour{\myFrameVec{w}{\nu}{}}{\myFrameVec{w}{\eta}{}}{\myFrameScalar{w}{\myFrameScalar{u}{c}{}}{}}{\myFrameScalar{w}{\myFrameScalar{v}{c}{}}{}} ,
\end{align}
with $\myFrameVec{w}{\nu}{}$ and $\myFrameVec{w}{\eta}{}$  as the process noise of the Fossen model and $\myFrameScalar{w}{\myFrameScalar{u}{c}{}}{}$ and $\myFrameScalar{w}{\myFrameScalar{v}{c}{}}{}$ as the process noise of the current model.    

For correction, three measurement models $\myFrameVec{h}{}{}(\myFrameVec{x}{}{})$ are used. The first corrects position $[n, e, d]^T$ with the USBL at a rate of 1\,Hz.
The second correct makes use of the pressure measurement of the depth sensor at a rate of 10\,Hz:
\begin{align}
    \myFrameScalar{z}{press}{} = k_p\cdot d + p_0 + v_{press},
\end{align}
with   
pressure per meter water column $k_p$,  
the atmospheric pressure $p_0$ and the pressure measurement noise $v_{press}$. 
The third corrects the attitude $[\phi, \theta, \psi]^T$  using the solution of the AHRS with 100\,Hz. 
We parameterize the process noise covariance matrix $\myFrameVec{Q}{}{} = \mathrm{diag}[\myFrameVec{Q}{\nu}{},\myFrameVec{Q}{\eta}{},\myFrameScalar{Q}{u_c}{},\myFrameScalar{Q}{v_c}{}]$ with four optimization parameters $\alpha_1,\dots,\alpha_4$ such that $\myFrameScalar{Q}{u_c}{} = \myFrameScalar{Q}{v_c}{} = \alpha_1$, $\myFrameVec{Q}{\nu}{} = \left[ \alpha_2 \mathbf{I}^{1\times3}, \alpha_2 \alpha_3 \mathbf{I}^{1 \times 3}  \right]^\mathrm{T}$ and $\myFrameVec{Q}{\eta}{} =  \alpha_4 \myFrameVec{Q}{\nu}{}$. The measurement covariance matrix $\mathbf{R}$ is assumed to be known.  

\subsection{Path planning}

Dubins curves are used  
to find the shortest path using the current estimated vehicle state and the next waypoint in the horizontal plane. 
The depth reference is generated by interpolating linearly between starting and goal depth (cf. e.g. \cite{Owen.2015}).
Dynamic vehicle constraints are considered by imposing a minimum turning radius. Waypoints are chosen such that a maximum glide path angle is not superseded. 
Additionally, a constant surge reference velocity is used. 
The Dubins curve-based path planner will be used in future work as the local planner within a graph based planning framework. In total, the path planner has two tunable parameters, the reference surge velocity $u_{\mathrm{ref}}$ and the Dubins path's radius $r_{\mathrm{plan}}$.

\subsection{Guidance}

As mentioned above, the nanoAUV is underactuated. For example, the sway and yaw components cannot be actuated independently from another.  
In order to resolve the underactuatedness, a line of sight guidance method with adaptive sideslip compensation \cite{FossenThorI..2015} is used. The main idea is to calculate a reference for the  
yaw angle, which reduces the  
horizontal deviation from the reference path using the kinematic relation 
\begin{equation} \label{eq:LOS_yaw}
    \psi_d =  \gamma_p + \mathrm{atan}(-\frac{h_e}{\Delta}) -\beta_{est} \text{.}
\end{equation}
The first term in Eq. \eqref{eq:LOS_yaw} $\gamma_p$  represents the path angle and the second term aims at reducing the horizontal crosstrack error $h_e$ (i.e. the shortest distance between the current AUV position and the reference path on the horizontal plane).
 
The 
side slip angle $\beta_{\mathrm{est}} = \mathrm{atan}(\frac{v}{u})$ occurs due to current induced drift.  
Here, $\beta_{\mathrm{est}}$ is calculated using the estimated velocities (cf. Sec. \ref{sec:stateEstimation}). This way, the state estimation also serves as a disturbance estimator. This approach is similar to estimating $\beta_{\mathrm{est}}$ directly within the guidance module \cite{FossenThorI..2015}. 
An analogue approach is used to calculate the pitch \hbox{reference $\theta_d$.}        

The quality of the estimation of $\beta$ is influenced by the parametrization of the state estimation. The main performance influencing tunable parameter within the guidance module is $\Delta$, which determines the aggressiveness of the guidance module.

\subsection{Control}

This section formulates a control law for the thrusters. 
We use the well known LQR method 
to synthesize a state feedback controller in order to track the surge velocity provided by the planner as well as pitch and yaw angles provided by the guidance. The controlled states are augmented with integral error states $u_i$, $\theta_i$ and $\psi_i$ resulting in an augmented state vector $\mathbf{x}_{\mathrm{fb}} =  [u, q, r,  \theta, \psi, u_i, \theta_i, \psi_i ]$.
The inclusion of integral error states is necessary in order to achieve offset free tracking in the presence of model plant mismatch. They are calculated by integration of the control error. 

The model for LQR synthesis is obtained by discretizing and linearizing Eq. \eqref{eq:Fossen_Mdl}. 
Here we choose a constant working point at $v=0$ and $ u=0.5 \, \mathrm{m/}\mathrm{s}$ with all other states set to zero. Potentially beneficial successive linearisation at different working points is not used to save computational resources.  

For the time discrete LQR, a state Feedback gain $\mathbf{K}$ is chosen such that the cost function 
\begin{equation} J = \sum_{k=1}^{\infty}  \mathbf{x}_{\mathrm{fb}}(k) \mathbf{Q}_\mathrm{LQR} \mathbf{x}_{\mathrm{fb}}(k)^T + \mathbf{u}_{\mathrm{lqr}}(k) \mathbf{R}_\mathrm{LQR} \mathbf{u}_{\mathrm{lqr}}(k)^T 
\end{equation} 
is minimized. The weighting matrices $\mathbf{Q}_\mathrm{LQR} = \mathrm{diag} \left[q_1, q_2, q_2, q_3, q_3, q_4, q_5,q_5\right] $ and $\mathbf{R}_\mathrm{LQR} = \mathbf{I}$ are diagonal matrices with unknown entries. The LQR control law follows as $\mathbf{u}_{\mathrm{lqr}} = [u_{\mathrm{surge}},u_{\mathrm{rl}},u_{\mathrm{ud}}] = -\mathbf{K} \cdot \mathbf{x}_{\mathrm{fb}}$. 
Additionally, a combination of a deadband and a hysteresis in order to avoid small high frequency thruster actuation is introduced.
In total, 6 tunable parameters $q_1,\dots,q_5$ and the deadband with $w_{\mathrm{db}}$ are considered.

\section{Simulation scenario} \label{sec:simulation_scenario}

\begin{figure} [h]
    \centering
    \includegraphics[width=0.9\linewidth]{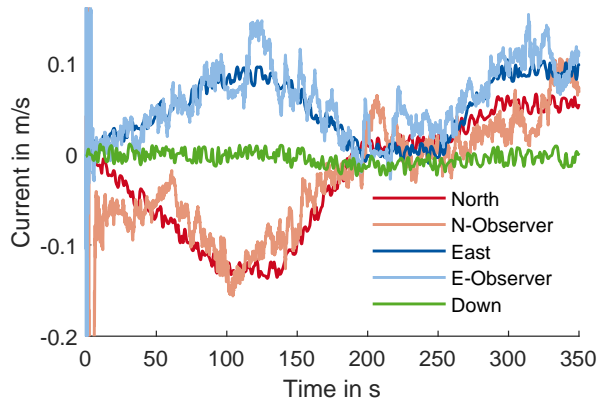}
    \caption{Randomly generated current profile: Ground truth in dark colors and the observers estimate of the north and east components in light color optimized for maximum controller tracking accuracy.}
    \label{fig:2sPlot}
\end{figure}

The plant model also relies on  Eq. \ref{eq:Fossen_Mdl}. However, in real world operation, model plant mismatch and disturbances deteriorate control and state estimation performance. Therefore,   
randomly drawn model plant mismatch is introduced in the simulation. The model used within the navigation filter and for controller synthesis deviates from the simulation in the hydrodynamic parameters (e.g. with a standard deviation of $10\%$ for the individual entries of $\mathbf{D})$, and the AUV mass. Additionally, the actuator lag is not considered in filter and controller. Furthermore, the static gain of the thruster transfer function is detuned with a standard deviation of $2.5\%$. 
 We introduce a randomly generated varying current as an additional disturbance (cf. Fig. \ref{fig:2sPlot}).  For the sensor simulation, USBL, pressure sensor and AHRS signals are perturbed with white Gaussian noise. Additionally, the USBL measurements are delayed by twice the sonic delay.  
 However,  
 the parameter tuning methodology is applicable to other, experimentally validated models, observed mismatches, and observed disturbances. 

The reference path is obtained by applying the path planner to a random sequence of way points. With the exception of Sec. \ref{resultsD} the same seed $s = 1$ for current, way points and model-plant mismatch is used. In these cases the simulated path has a length of $210 \, \text{m}$.  Except for Sec. \ref{resultsC} %and \ref{resultsD} 
the following arbitrary energy cost function
\begin{equation} \label{eq:ActcostFun}
    J(\boldsymbol{a}|s) = T_{\mathrm{end}}(\boldsymbol{a}|s) + \sum_{i = \{ \mathrm{surge},\mathrm{rl},\mathrm{ud}\} } \bar{J}_i(\boldsymbol{a}| s)
\end{equation}
 with 
\begin{align} \label{eq:thrusterEnergy}
    \bar{J}_i(\boldsymbol{a}) &= \int_{t= 0}^{T_{\mathrm{end}}(\boldsymbol{a|s})} |\mathrm{sgn}(u_i(t|\boldsymbol{a},s))|  \cdot \dots \\ 
    &\left(0.025 + |u_i(t|\boldsymbol{a},s) + u_i(t|\boldsymbol{a},s)^{1.5}| \right) \mathrm{dt}  \nonumber
\end{align} is chosen. The time which the AUV needs to reach the final waypoint for a given parametrization $\boldsymbol{a}$ and seed $s$ is denoted as $T_{\mathrm{end}}(\boldsymbol{a},s)$, therefore the first term in Eq. \eqref{eq:ActcostFun} represents consumers with constant power demand such as sensors. The energy consumed by each thruster is calculated via Eq. \eqref{eq:thrusterEnergy}. Power consumption is only zero if the control input is exactly zero. %Otherwise a minimum energy consumption is assumed. 
The maximum observed deviation from the reference path $g(\boldsymbol{a},s)$ is constrained by $g_\mathrm{max} = 1.5 \, \mathrm{m}$. In total, one execution of the simulation takes around 30 seconds\footnote{All experiments were performed on a notebook with an Intel Core i7-10510U Processor @2.3GHZ and 16 GB Ram.}.

\section{Optimization of the GNC system} \label{sec:results}

In the subsequent section, we apply the tuning method to the presented GNC system. It is examined whether the whole GNC system needs to be tuned at once (cf. Sec. \ref{resultsA}), different accuracy requirements (cf. Sec. \ref{resultsB}) and different cost functions (cf. Sec. \ref{resultsC}) require different parametrizations, and (cf. Sec. \ref{resultsD} robustness can be increased. 
Each test case is ran 5 times with $45 \cdot d$ objective function evaluations (simulations) each, where $d$ is the number of tunable parameters. For the overall GNC system with 13 parameter one optimization takes around $6.5 \, \mathrm{h}$. The tuning parameters for the $\mathbf{Q}$ matrices of state estimation and controller are optimized in the log domain.

\subsection{Joint vs. individual Optimization} \label{resultsA}

Table \ref{tab:ResultsIndvCombi} compares the combined optimization with the results if only the parameters of the respective GNC submodules are optimized. The default parametrization is hand tuned and does not make use of the deadband. Results indicate that energy consumption can significantly be decreased if the GNC system is optimized as a whole instead of optimizing only one individual component. Additionally, it can be seen that the optimizer finds a good solution reasonably consistent since the deviations of the best and worst run are limited. 

\begin{table}[h] 
	\caption{Individual vs. joint optimization - Best (Worst) of five optimization runs.}
	\centering
	\begin{tabular}{p{1.5cm}| >{\raggedleft}p{0.9cm} >{\raggedleft}p{0.9cm} >{\raggedleft}p{0.9cm} >{\raggedleft}p{0.9cm} >{\raggedleft}p{0.9cm} }
		\toprule
		  Optimized \newline system & Plan.  \& \newline Guid.          & Control                & Filter               & Combi.    & Default      \tabularnewline
		\midrule
		    No. of Params $d$ & $3$                       &  $6$              &  $4$            & $13$  & -	 
		   \tabularnewline
		   \rule{0pt}{3ex}
		   Energy Consumption $J$ & $\, 106.6$\newline  $\quad (111.9)$                       &  $\, 118.6$ \newline $\quad (121.1)$              &  $\,135.3$ \newline  $\quad (136.4)$            & $\quad 87.6$ \newline  $\quad (94.2)$  & $\quad 137.1$	   
		    \tabularnewline 
		   \rule{0pt}{3ex}  
		   maximum \newline Deviation $g$ & $\quad 1.46$\newline $\quad (1.43)$                       &  $\quad 1.49$ \newline $\quad (1.06)$              &  $\quad 1.36$  \newline $\quad (1.22)$            & $\quad 1.36$  \newline $\quad (1.46)$              & $\quad 1.11$	   
		    \tabularnewline 
		\bottomrule
	\end{tabular}
	\label{tab:ResultsIndvCombi}
\end{table}

\subsection{Energy consumption vs. path deviation} \label{resultsB}

In this section, we use the joint optimization of all GNC parameters to examine the conflict of goals between tracking performance and energy consumption. Three different accuracy scenarios are defined. They differ in the maximum allowed planning radius $r_\mathrm{plan}$ and path deviation requirements $g_\mathrm{max}$. In the maximum accuracy scenario, the maximum observed path deviation is minimized in an unconstrained optimization without considering energy consumption. 
Table \ref{tab:ResultsAcc} show the results. The conflict of goals becomes apparent especially when comparing the maximum and medium accuracy scenarios. The low accuracy scenario only slightly decreases energy consumption although the tracking performance is decreased significantly. 
With decreasing accuracy requirements, we observe an increase in $u_{\mathrm{ref}}$, $r_{\mathrm{plan}}$ and  $\Delta$ which meets expectations. However, for other parameters such as $w_{\mathrm{db}}$ and $q_4$ the trend is not as clear. Additionally,
UKF parameters are tuned towards higher position accuracy if better tracking performance is required.  
In contrast, smoothness in the current estimation is surprisingly not essential for maximum tracking performance (cf. Fig. \ref{fig:2sPlot}).
The best reached tracking accuracy is deemed satisfactory. For collision avoidance, different sensors (e.g. echosounders) and guidance strategies will be used. 

\begin{table}[h] 
	\caption{Different accuracy requirements - Best of five runs.}
	\centering
	\begin{tabular}{l | r r r }
		\toprule
		  Scenario          & Max. Acc.  & Med. Acc.               & Low Acc.       \\
		\midrule
		Energy consumption $J$             &  $121.1$ 	            &  $87.6$           &  $85.2$               \\ 
		   Max. deviation $g$             &  $0.82$              &  $1.36$            & $2.17$  	   \\ 
		
		Max. pos. estimation error              &  $0.67$ 	            &  $1.10$           &  $1.13$               \\
				\midrule
		Max. allowed $r_\mathrm{plan}$ &  $5$ 	            &  $10$           &  $15$\\  
	    Deviation lim. $g_\mathrm{max}$            &  min! 	            &  $1.5$           &  $3$\\ 
		\midrule
		Optimized $u_{\mathrm{ref}}$             &  $0.47$ 	            &  $0.81$           &  $0.89$               \\
		Optimized $r_{\mathrm{plan}}$             &  $4.98$ 	            &  $5.74$           &  $7.01$               \\ 
		Optimized $\Delta$             &  $4.0$ 	            &  $59.0$           &  $89.3$               \\ 
		Optimized $\mathrm{log10} \, (q_4)$    & $-3.08$ & $-3.25$ & $-2.88$ \\
        Optimized $w_\mathrm{db}$      & $0.06$ & $0.13$ & $0.08$ \\
		\bottomrule
	\end{tabular}
	\label{tab:ResultsAcc}
\end{table}

\subsection{Varying cost function} \label{resultsC}
\begin{figure} [h]
    \centering
    \includegraphics[width=0.9\linewidth]{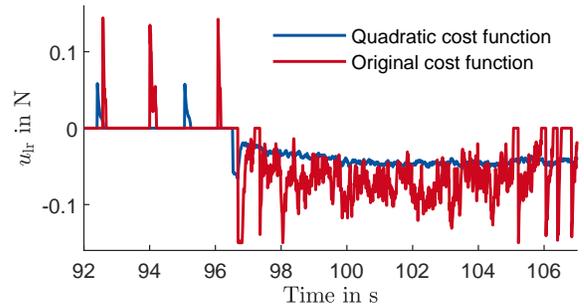}
    \caption{Control signal in lateral direction optimized for different power consumption functions. Straight line tracking followed by a right hand curve starting at $96.5 \, \mathrm{s}$.}
    \label{fig:uPlot}
\end{figure}

The method may also be used to automatically adapt the GNC system for different actuator designs. In order to mimic different actuator designs, we modify the thruster energy function Eq. \eqref{eq:thrusterEnergy} to a purely quadratic one. Fig. \ref{fig:uPlot} shows the resulting different actuator control signals. Both policies exhibit pulses when following a straight line and a more continuous control input when cornering. However, the optimal time domain behavior changes significantly if different energy consumption characteristics are used. Additionally, the optimized reference velocities $u_{\mathrm{ref}}$ is reduced from $0.81 \mathrm{m/}\mathrm{s}$ (original) to $0.48 \mathrm{m/}\mathrm{s}$ (quadratic).   

\subsection{Optimization for robustness} \label{resultsD}

Up to now, we only optimized for one fixed model plant mismatch and disturbance trajectory $s =  1$ by using the obj. fun. $J(a|s=1)$. In practice, the exact model plant mismatch is likely not known a-priori. When validating the Low. Acc. scenario (cf. Sec. \ref{resultsB}) on 25 different randomly drawn seeds $s$, eleven violate the path deviation contraint. This indicates that over-fitting is a significant issue, if only one seed is considered for parameter optimization. To tackle this, the same five randomly drawn seeds are evaluated at each objective function query step (cf. Step 6 in Algo. 1). Afterwards, energy consumption is averaged over the five runs \hbox{$\hat{J}(a) = 1/5 \sum_{s = 1}^5 J(a|s)$} and the maximum path deviation is taken $\hat{g}(a) = \max_{s \in \{ 1, ..., 5 \} } g(a|s) $. As a result optimization takes five times as long. Apart from that, the optimization problem Eq. \eqref{eq:generalOptProb} is not altered. Results show, that robustness is increased to only one infeasible validation sample at only $\sim 5\%$ more energy consumption on average.

\section{CONCLUSIONS} \label{sec:conclusions}

In this contribution, we have demonstrated the versatile 
applicability of a BO-based automatic parameter tuning methodology to the holistic optimization of the GNC system of an underactuated AUV. The method can be applied to arbitrary high-level control objectives, vehicles and operational scenarios. As expected, only overall tuning yields the lowest energy consumption. The systematic optimization-based approach, results in a difference of minus $18 \%$ compared to the best individual optimization and minus $36 \%$ compared to the hand-tuned default parametrization.
The method is able to tackle various aspects of the GNC system design procedure such as different actuator designs and robustness. Robustness can be increased drastically by optimizing on various environmental conditions. 
Furthermore, the method was used to automatically generate different parametrizations for different accuracy requirements. Results indicate, that different requirements can be met with the same GNC structure. This suggests that during operation, different precomputed parameter sets should be used depending on the varying mission requirements. 

In future work, the optimization will be extended by an experimentally validated AUV model. Furthermore, it will be investigated in which way the optimized parameters can improve the performance in real experiments.

\bibliography{IEEEabrv,main}
\addtolength{\textheight}{-6cm}

\end{document}